# THE EFFECT OF COMMUNICATION AND SYNCHRONIZATION ON AMDAHL'S LAW IN MULTICORE SYSTEMS

L. Yavits, A. Morad, R. Ginosar

**Abstract**—This work analyses the effects of sequential-to-parallel synchronization and inter-core communication on multicore performance, speedup and scaling. A modification of Amdahl's law is formulated, to reflect the finding that parallel speedup is lower than originally predicted, due to these effects. In applications with high inter-core communication requirements, the workload should be executed on a small number of cores, and applications of high sequential-to-parallel synchronization requirements may better be executed by the sequential core, even when $f$, the Amdahl's fraction of parallelization, is very close to 1. To improve the scalability and performance speedup of a multicore, it is as important to address the synchronization and connectivity intensities of parallel algorithms as their parallelization factor.

———————————— ◆ ————————————

## 1 INTRODUCTION AND RELATED WORK

CONSIDER a multicore comprising a general purpose core responsible for the sequential fraction of the code (the *sequential core*) and a number of parallel cores that execute the parallelizable fraction of the program. Parallelization incurs data exchange between the sequential core and the parallel cores at the beginning and the end of each parallel section of a program (*sequential-to-parallel synchronization*), and the data exchange among the parallel cores during parallel execution (*inter-core communication*). This paper investigates the effect of these synchronization and communication elements on multicore performance, speedup and scaling.

Many multicore performance models tend to ignore the impact of sequential-to-parallel synchronization and inter-core communication on performance and power. We propose an analytic model that accounts for these effects. Based on this model, we reach a number of conclusions that affect multicore design. First, we show that impact of sequential-to-parallel synchronization and inter-core communication on performance becomes more significant with the level of parallelism. Second, a smaller number of larger cores tends to be more efficient performance-wise than a larger number of smaller cores as implied by Amdahl's law, and this effect becomes more predominant with growing parallelism. Third, for low arithmetic intensity [22] tasks, parallel execution may result in lower speedup, or even no speedup relative to Amdahl's law [1]. We also arrive at a number of conclusions as to an optimal task scheduling: for tasks with high inter-core communication requirements, the workload should be assigned to a small number of cores, even if the parallelization fraction $f$ is very close to 1. For low arithmetic intensity tasks, the workload had better be assigned to the sequential core, even if the parallelization fraction $f$ is very close to 1. Both conclusions stand in contradiction to Amdahl's law, which implies that the higher $f$ is, the more cores should be used.

Extending Amdahl's law in the multicore era and modeling multicore performance has been thoroughly studied. Hill and Marty augmented Amdahl's law with a corollary to multicore architecture by constructing a model for multicore performance and speedup [2]. Cassidy *et al.* [5] [6][7] extended the model by considering also cache area and power, and suggested an optimization framework. Eyerman and Eeckhout introduced a model that accounted for critical sections of the parallelizable fraction of a program [7]. Gunther *et al.* considered the effects of resource sharing and limitations on such models [10].

In this paper, we analyze the work of Hill and Marty, Cassidy et al., Eyerman and Eeckhout and Gunther et al., introduce a novel model that considers the effects of sequential-to-parallel synchronization and inter-core communication, and compare all five models in a unified framework.

Other studies also discussed multicore scalability and performance. Oh et al. investigated the tradeoff between the number of CMP cores and cache architecture, including cache size, L2 and L3 cache effects, shared vs. private and hybrid caches, and uniform vs. non-uniform caches [4]. Chung et al. extended Hill and Marty's model by considering the constraints of off-chip bandwidth, and by introducing accelerators achieving better performance / power ratio [9]. Sun and Chen [18] took a different approach to multicore performance modeling, based on Sun

————————————————

- *Leonid Yavits (\*), E-mail: yavits@tx.technion.ac.il.*
- *Amir Morad (\*), E-mail: amirm@tx.technion.ac.il.*
- *Ran Ginosar (\*), E-mail: ran@ee.technion.ac.il.*
- *(\*) Authors are with the Department of Electrical Engineering, Technion-Israel Institute of Technology, Haifa 32000, Israel.*

and Ni [23] rather than Amdahl's law. Rogers et al. studied the limitations of multicore scaling imposed by memory bandwidth constraints and created an analytic model for memory traffic [11]. Loh augmented the Hill and Marty's model by accounting for the cost (area) of the "uncore", namely the parts of the chip outside the core [12]. Synchronization and communication overhead was originally studied by Flatt and Kennedy [32]. Morad *et al.* [17] incorporated this overhead into the multicore performance model.

The rest of this paper is organized as follows. Sect. 2 starts by discussing factors that limit speedup, continues to describe Hill and Marty's, Cassidy *et al.*, Eyerman and Eckhout's and Gunther *et al.* models, and introduces our models for symmetric and asymmetric multicore architectures. Sect. 3 describes simulations that validate the models. Sect. 4 offers a discussion of the results, and Sect. 5 presents the summary.

## 2 MULTICORE PERFORMANCE MODEL

The section starts by presenting four existing models in a unified framework and proceeds to introduce a novel model for symmetric and asymmetric multicore architectures.

### 2.1 Beyond Amdahl's law – speedup limiting factors

Our novel multicore performance model accounts for the effects of inter-core communication and data synchronization, as follows. Consider the asymmetric multicore model shown in Figure 1, the symmetric multicore model in Figure 2 and the sequential / parallel execution model of Figure 3. Examples of asymmetric (even heterogeneous) systems include a CPU with its own memory, accompanied by a GPU with graphics memory, and the CELL processor [27], and are analyzed in [17]. Tiled architectures [28] [29] [30] are examples of symmetric systems.

Sequential code segments are executed on either the sequential core (in the asymmetric architecture) or one of the cores (in the symmetric architecture), and parallel segments are executed on the parallel multicore. In the asymmetric multicore, these two phases employ two different memories: the sequential memory and the multiple memories of the multicore. In the symmetric architecture, the sequential phase employs the local memory of one core, while the parallel phase again employs the multiple memories of the multicore. In between parallel and sequential segments, data are synchronized, namely transferred between the two memories, as indicated by the double arrows in the two figures. That transfer is named *sequential to parallel synchronization*.

During parallel execution, the multiple cores may exchange data. That *inter-core communication* comprises the second limiting effect studied in this paper.

Based on Amdahl's law [1], the execution time of such multicore system can be written as follows:

$$T_1 = (1-f)T_1 + fT_1 \qquad (1)$$

where $T_1$ is the sequential execution time of the program (on the sequential core) and $f$ is the parallelizable fraction of the program.

The cores of the parallel multicore are connected to each other and to the sequential core (and its memory) through an interconnection network. The access time from the parallel multicore to the memory of the sequential core depends on the number of parallel cores $n$ as well as on the interconnection network configuration. While typical network configurations incur communication distance and latency of $O(\log n)$ hops, we consider $O(\sqrt{n})$ models such as 2D mesh networks for mathematical simplicity. This cost model applies to both effects, sequential-to-parallel synchronization and inter-core communication.

#### 2.1.1 Sequential-to-Parallel Data Synchronization

An example of sequential-to-parallel data synchronization is the CudaMemCpy procedure of the NVIDIA CUDA parallel architecture [15].

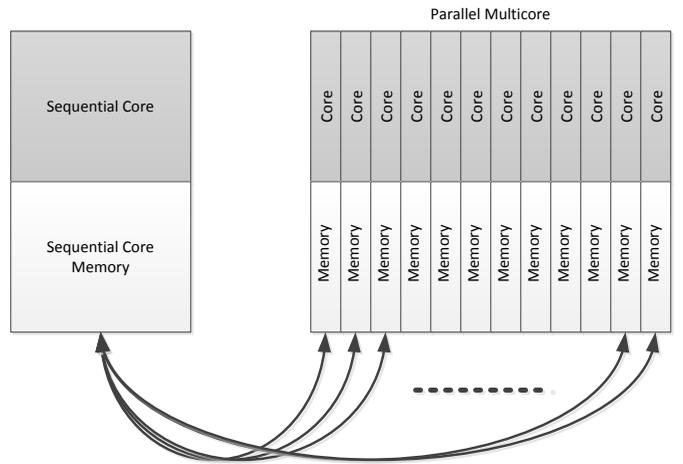

Figure 1. Asymmetric multicore model

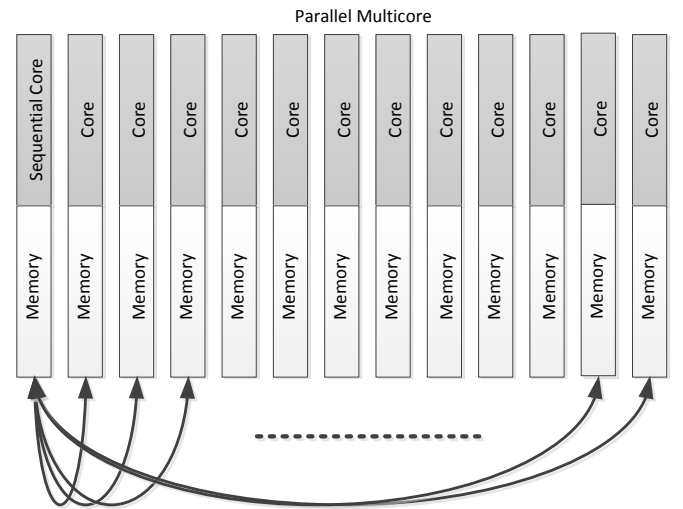

Figure 2. Symmetric multicore model

While often discounted in multicore architecture analysis, data synchronization between the sequential and parallel portions of a program may take a significant fraction of total runtime. Each time a parallelizable segment of a program is scheduled to execute on parallel cores, these cores need to download the input data from the sequential core memory into their distributed local memory (through a bandwidth-limited interface), execute the parallel code and then upload the output data or results to the sequential core memory. The effect of data synchronization between the sequential and parallel portions of a program is common to all multicore cache models, including private, shared and hybrid cache. If the data needed for parallel execution is located in a higher level memory (shared by the sequential core and the parallel cores) rather than in the private memory of sequential core, it still needs to be transferred to the individual memories of parallel cores prior to the execution of the parallel session.

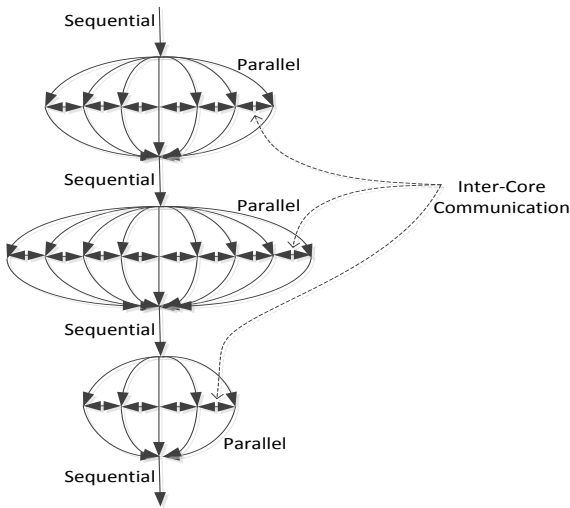

Figure 3: Sequential / Parallel Execution Model

Sequential-to-parallel synchronization is affected by *arithmetic intensity*, which is defined as ratio of arithmetic operations to memory traffic [22]. Following [26], Figure 4 shows the distribution of arithmetic intensity over a wide range of parallel algorithms, spanning from $O(1)$ through $O(\log N)$ to $O(N)$.

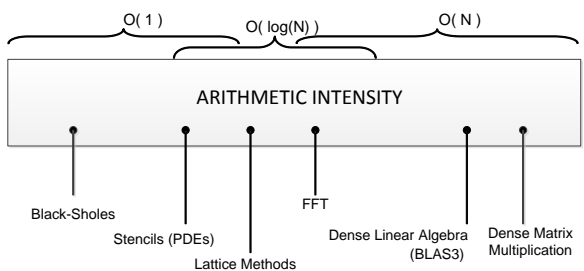

Figure 4: Arithmetic Intensity [26].

We employ three examples of algorithms which span a wide spectrum of arithmetic intensity to demonstrate its effect on sequential-to-parallel data synchronization: Black and Scholes option pricing [13][31], FFT and dense matrix multiplication.

- For Black and Scholes pricing of $N$ options, the task size, and hence also the complexity of sequential-to-parallel data synchronization, is $O(N)$, and the computational complexity is also $O(N)$. Thus, arithmetic intensity is $O(1)$.
- For $N$ point FFT, the task size and the sequential-to-parallel data synchronization complexity is $O(N)$, while the computational complexity is $O(N \log N)$. Arithmetic intensity is $O(\log N)$.
- For $N \times N$ dense matrix multiplication, the task size and the sequential-to-parallel data synchronization complexity is $O(N^2)$, while the computational complexity is $O(N^3)$. Arithmetic intensity is hence $O(N)$.

Consequently, we define:

**Definition 1:** *Synchronization intensity* is the ratio of number of data elements transferred during sequential-to-parallel data synchronization to the number of arithmetic operations. Alternatively, it is the ratio of the time it takes to complete the sequential-to-parallel data synchronization in a serial manner, to the sequential execution time of a program, as defined in (1).

Note that synchronization intensity is inversely proportional to the arithmetic intensity.

### 2.1.2 Inter-core Data Communication

Inter-core data communication is indicated by horizontal arrows in the parallel sessions of Figure 3. In order to quantitatively address the effect of inter-core communication, we define:

**Definition 2:** *Connectivity intensity* is the ratio of number of data elements transferred during inter-core communication to the number of arithmetic operations. Alternatively, it is the ratio of the time it takes to complete the inter-core communication in a serial manner, to the sequential execution time of a program as defined in (1).

The effect of inter-core communication on multicore performance can be demonstrated using the same examples as above:

- Black and Scholes option pricing requires no interaction between separate option calculations, therefore with serial computing time of $O(N)$, its connectivity intensity is zero (Black and Scholes option pricing is an example of an 'embarrassingly parallel' task).
- For $N$ point FFT, serial computing time is $O(N \log N)$. For $N$ core parallel implementation, computing time is reduced to $O(\log N)$ steps. But after each computing step, $N$ intermediate results need to be exchanged among the cores. Therefore the connectivity intensity is $(N \cdot O(\log N)) / (O(N \log N)) = 1$
- For $N \times N$ matrix multiplication, computational complexity and serial execution time are $O(N^3)$. One possible implementation uses $n = N \times N$ cores, yielding parallel execution time of $O(N)$, with $O(N^2)$ data elements being shifted every step [19]. Therefore connectivity intensity is $(N^2 \cdot O(N))/(O(N^3)) = 1$. An alternative implementation eliminates the need for inter-core communication by storing the relevant row and column of two multiplicand matrices in each processing core. For such implementation, the connectivity intensity is zero.

As the above examples show, the inter-core connectivity time and connectivity intensity strongly depend on the specifics of each workload as well as actual implementation of the architecture, for example the number of cores.

Synchronization and connectivity intensities can be best presented on a two-dimensional chart (Figure 5). The upper left corner marks the lowest connectivity and synchronization intensities. The algorithms located in this corner are best suited for parallel implementation. The lowest connectivity intensity is 0, which is typical for "embarrassingly parallel" tasks where there are no inter-core data dependencies. The lowest synchronization intensity we consider in our analysis is $1/N$ which is characteristic of dense linear algebra algorithms such as dense matrix multiplication. The lower right corner marks the highest connectivity and synchronization intensities. The algorithms located in that corner are least suited for parallel implementation, even if they allow high parallelization, since the synchronization and communication overheads significantly dampen the parallel speedup as we show in Sections 2.3 and 2.4 below. The highest connectivity intensity we consider in our analysis is 1 (meaning that the sequential computation and sequential inter-core communication times are of the same asymptotic order). This figure is typical for FFT. The highest synchronization intensity we consider in our analysis is 1 (meaning that the sequential computation and sequential–to-parallel synchronization times are of the same asymptotic order). This figure is typical for Black-Scholes option pricing. The relative positions of these three examples on the two-dimensional synchronization / communication intensity diagram are presented in Figure 5.

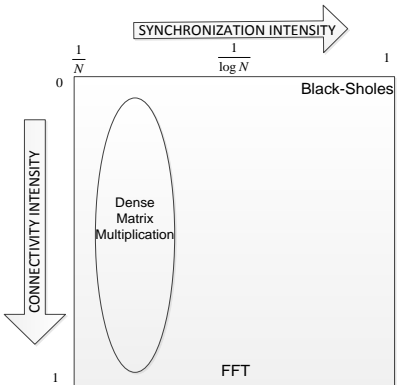

Figure 5: Synchronization and Connectivity Intensity

## 2.2 Comparative performance analysis

In this section, we compare some of the existing multicore performance models using Hill and Marty's model [2] as a reference framework.

### 2.2.1 Amdahl's law in multicore era

Hill and Marty [2] envision a symmetric or asymmetric multicore comprising base core equivalent (BCE) cores and cores that are $r$ times larger. The symmetric multicore consists of $n/r$ $r$-sized cores and the asymmetric multicore comprises one $r$-sized core and $n - r$ 1BCE cores. The speedup of the multicore relative to the BCE performance (which is assumed unity) can be expressed as follows:

$$Speedup_{mc} = \frac{T_{seq}(BCE)}{T_{mc}} = \frac{T_1(r)}{T_{mc}} \cdot \frac{T_{seq}(BCE)}{T_1(r)} \quad (2)$$
$$= \frac{T_1(r)}{T_{mc}} \cdot Perf_{seq}(r)$$

where $T_{seq}(BCE)$ is the sequential execution time of the program on 1BCE core, $T_1(r)$ is the sequential execution time of the program on $r$-sized core as defined in (1), $T_{mc}$ is the parallel execution time of the program on a multicore and $Perf_{seq}(r)$ is the performance of a sequential $r$-sized core.

This speedup function can also be generalized as follows (assuming the $r$-sized sequential core is used during parallel execution as well):

$$Speedup_{mc} = \quad (3)$$
$$= \frac{1}{\frac{1-f}{Perf_{seq}(r)} + \frac{f}{Perf_{seq}(r) + Perf_{par}(n,r)}}$$

where $Perf_{par}(n,r)$ is as follows:

$$Perf_{par}(n,r) = \quad (4)$$
$$= \begin{cases} \frac{n-r}{r}\sqrt{r} \text{ for symmetric multicore} \\ n-r \text{ for asymmetric multicore} \end{cases}$$

### 2.2.2 Linking multiprocessor performance to CPU and cache area

Cassidy et al. [5] [6][7] created a comprehensive analytical model that connects the performance of symmetric multicore to the number of cores, to CPU and $L_2$ cache area and to power. Cassidy et al. model accounts for $L_2$ cache misses and external shared memory access, and incorporates the area cost of $L_2$ cache and fixed area functions into overall performance cost function. This model though addresses neither the interference between cores nor cache coherence [7]. We can present Cassidy et al. speedup (based on (55) in [7]) as follows:

$$Speedup_{sym} = \frac{J_D^{Po}}{J_D} = \frac{1}{J_D} \quad (5)$$

And ((42) in [7]):

$$J_D = \left(1 - f + \frac{f}{n_c}\right) \cdot \left[G_0\left(\beta A_P^{-\frac{1}{2}}\right)\right. \quad (6)$$
$$+ (1 - G_0)\left(1 - kA_{L2}^{-\frac{1}{2}}\right)D_1$$
$$\left. + (1 - G_0)\left(kA_{L2}^{-\frac{1}{2}}\right)D_2\right]$$

where $A_p$ is the processor area, $A_{L2}$ is the $L_2$ cache area, $G_0$ is the fraction of instructions that require no memory access beyond $L_1$ cache, $\beta$ and $k$ are constant parameters, $D_1$ and $D_2$ are the access times of $L_2$ cache and external memory, respectively, as defined in Cassidy et al. [6] [7][6] . $D_1$ and $D_2$ are also constants in [6] [7][6] . In order to adapt (5) to our framework, we express the area variables using $n$ and $r$ as follows:

$$A_p = f_P \cdot r$$
$$A_{L2} = f_C \cdot r$$
$$n_c = \frac{n}{r} \quad (7)$$

where $n_c$ is the number of cores, $f_P$ is the fraction of the core occupied by processor, $f_C = 1 - f_P$ is the fraction of the core occupied by $L_2$ cache. It should be noted that Cassidy et al. created an optimization framework where $A_p$, $A_{L2}$ and $n_c$ are optimized so as to minimize the cost function $J_D$ under the total area constraint. We assume constant $f_P$ and $f_C$ which potentially leads to a suboptimal solution (in terms of $J_D$ minimization). This assumption is acceptable for the purpose of our comparative analysis.

### 2.2.3 Modeling critical sections

Eyerman and Eeckhout [7] concluded that parallel performance is not only limited by sequential code (as suggested by Amdahl's law) but is also fundamentally limited by synchronization through critical sections. They modeled two cases of critical section execution: execution time determined by average thread and execution time determined by slowest thread, as follows (for symmetric multicore):

$$T_{avg} = f_{par,cs} \cdot p_{cnt} \cdot p_{cs}$$
$$+ \frac{f_{par,cs} \cdot (1 - p_{cnt} \cdot p_{cs}) + f_{par,ncs}}{n_c}$$

and
$$T_{slw} = f_{par,cs} \cdot p_{cnt} \quad (8)$$
$$+ \frac{f_{par,cs} \cdot (1 - p_{cnt}) + f_{par,ncs}}{2n_c}$$

where $n_c = n/r$; the second summand is assumed to be accelerated by parallel execution and the first summand is performed sequentially.

To adapt Eyerman and Eeckhout's model to our framework, we rewrite the accelerated $T_{avg\,sym}$ and $T_{slw\,sym}$ as follows:

$$T_{avg\,sym} = \frac{f_{par,cs} \cdot p_{cnt} \cdot p_{cs}}{\sqrt{r}}$$
$$+ \frac{r \cdot (f_{par,cs} \cdot (1 - p_{cnt} \cdot p_{cs}) + f_{par,ncs})}{n\sqrt{r}}$$

and

$$T_{slw\,sym} \quad (9)$$
$$= \frac{f_{par,cs} \cdot p_{cnt}}{\sqrt{r}} + \frac{r \cdot (f_{par,cs} \cdot (1 - p_{cnt}) + f_{par,ncs})}{2n\sqrt{r}}$$

where $r$ is the core size, $n/r = n_c$ is the number of cores and $\sqrt{r}$ is the core performance. The speedup for symmetric multicore (relative to a single BCE core as in [2]) can then be written as follows:

$$Speedup_{sym} = \frac{1}{\frac{f_{seq}}{\sqrt{r}} + \max(T_{avg\,sym}, T_{slw\,sym})} \quad (10)$$

The speedup for asymmetric multicore (relative to a single BCE core as in [2]) can be written as follows:

$$Speedup_{asym} = \frac{1}{\frac{f_{seq}}{\sqrt{r}} + \max(T_{avg\,asym}, T_{slw\,asym})} \quad (11)$$

where

$$T_{avg\,asym} = \frac{f_{par,cs} \cdot p_{cnt} \cdot p_{cs}}{\sqrt{r}}$$
$$+ \frac{f_{par,cs} \cdot (1 - p_{cnt} \cdot p_{cs}) + f_{par,ncs}}{\sqrt{r} + n - r}$$

and

$$T_{slw\,asym} = \frac{f_{par,cs} \cdot p_{cnt}}{\sqrt{r}} + \frac{f_{par,cs} \cdot (1 - p_{cnt}) + f_{par,ncs}}{2(\sqrt{r} + n - r)}$$

where $\sqrt{r}$ is the performance of the sequential core and $f_{seq}, f_{par,cs}, f_{par,ncs}, p_{cnt}, p_{cs}$ are sequential portion of the program, parallelizable portion of the program containing critical sections, parallelizable portion of the program that requires no synchronization, contention probability and critical section probability respectively as defined in [7].

Eyerman and Eeckhout work is preceded by research of M. Suleman et al [21], who discussed the "serialization" effect of critical sections of parallel threads and its impact on parallel speedup.

### 2.2.4 Universal Scalability Model

Gunther et al. [10] universal scalability model accounts for waiting for access to shared resources, retrograde scaling (latency due to exchange of data between caches) and resource limitations. According to Gunther et al. model, parallel speedup is as follows:

$$Speedup_{par}(n) = \frac{n}{1 + \alpha(n-1) + \beta n(n-1)} \quad (12)$$

Where $n$ is the number of parallel cores as well as the scaled size of the workload, and $\alpha$ and $\beta$ are scalability parameters. To adapt Gunther et al. model to our frame-work, we need to substitute (12) into (3), and thus we receive the symmetrical multicore speedup by Gunther et al.:

$$Speedup_{sym} = \frac{1}{\frac{1-f}{\sqrt{r}} + \frac{f}{\sqrt{r} \cdot Speedup_{par}(\frac{n}{r})}} \quad (13)$$

### 2.3 Symmetric multicore performance model

Taking into account the factors reviewed above, we can write the multicore execution time as follows:

$$T_{mc}(n_c) = (1-f)T_1 + \frac{fT_1}{n_c} + \frac{T_C}{n_c} + T_S \quad (14)$$

where

$n_c$ is the number of cores;

$f$ is the parallelizable fraction of the program as defined in (1);

$T_1$ is the sequential execution time of the program as defined in (1);

$T_C$ is the time it takes to complete inter-core communication in a sequential manner, and is a function of the number of cores as suggested above. Since a multicore processor normally features a parallel inter-core communication network, such as 2D mesh, we can assume that the inter-core communication time scales in a manner similar to the execution time.

$T_S$ is time to complete the sequential-to-parallel data synchronization in a sequential manner. It depends on the data size $N$; since sequential-to-parallel synchronization involves access to a shared resource, $T_S$ might also depend on the number of cores [4][11]. This is especially the case when the task size is scaled down to the multicore size, i.e. when $N = n_c$. For this reason, we can assume that the sequential-to-parallel data synchronization time does not scale the same way as do the execution and the inter-core communication times.

The speedup function of the symmetric multicore (relative to one single BCE core as defined in [2]) can be written as

$$Speedup_{sym} = \frac{Perf_{seq}(r) \cdot T_1}{T_{mc}} \quad (15)$$

$$= \frac{Perf_{seq}(r)}{(1-f) + \frac{f}{n_c} + \frac{T_C}{n_c T_1} + \frac{T_S}{T_1}}$$

where the number of cores $n_c = n/r$.

Substituting $Perf_{seq}(r) = \sqrt{r}$, the connectivity intensity $f_1 = T_C / T_1$ and the synchronization intensity $f_2 = T_S / T_1$, we can rewrite the symmetric multicore speedup as follows:

$$Speedup_{sym} = \frac{\sqrt{r}}{(1-f) + \frac{f}{n_c} + \frac{f_1}{n_c} + f_2} \quad (16)$$

Substituting $f_1 = f_2 = 0$ in (16) yields Hill and Marty symmetric multicore model [2] or Cassidy's simplified model as per (56) in [7].

Figure 6 shows the speedup of the symmetric multicore using our model, Hill and Marty's model [2], Cassidy et al. [7] model, Eyerman and Eeckhout (EE) model [7] and Gunther et al. model [10].

For modeling purposes we assume the total BCE count $n = 256$ and $f = 0.5, 0.95, 0.99$ and $0.999$. We also assume the following parameters, in order to produce comparable speedup values of the different models shown in Figure 6:

For Cassidy et al. model, $f_P = 0.66$ and $f_C = 0.34$ while the rest of the parameters are as defined in [6].

For Gunther et al. model, α=β=0.001.

For Eyerman and Eeckhout model, $f_{seq} = f$, $f_{par,ncs} = 0.9 \cdot (1 - f_{seq})$, $f_{par,cs} = 0.1 \cdot (1 - f_{seq})$, $p_{cnt} = 0.1, p_{cs} = 0.1$.

For our model, we assume $f_1 = O(n^{0.5}) = 0.001 \cdot n^{0.5}$ and $f_2 = 0.01$. We address the relation of multicore performance to $f_1$ and $f_2$ in greater details in Sect. 4.

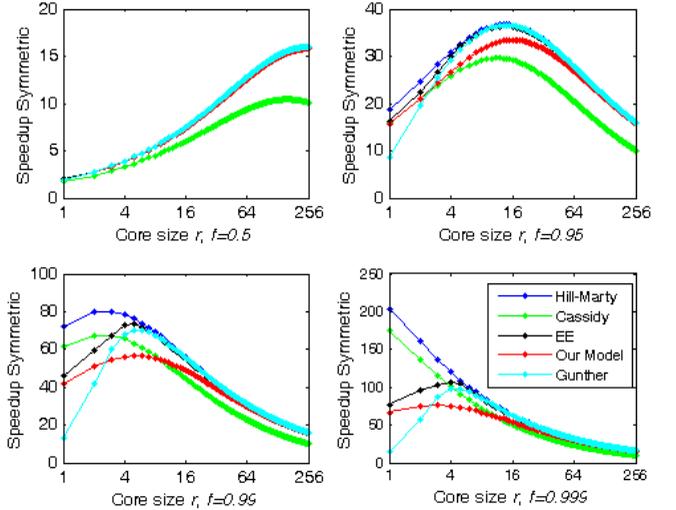

Figure 6. The speedup of the symmetric multicore vs. core size $r$

When parallelism is low (f=0.5), all models behave similarly (except for Cassidy et al, because a fraction of BCE resource is explicitly assumed to be allocated to cache, thus making the effective core count, and therefore the speedup, lower). The higher the parallelism, the more apparent is the difference between the models. According to Hill and Marty and Cassidy et al. models, optimal speedup is achieved when the core size is small (eventually reaching $r$=1BCE for $f = 0.999$) and the number of cores is large (eventually reaching $n_c = n$ for $f = 0.999$). Eyerman and Eeckhout, Gunther et al. and our model reach a different conclusion. The optimal core size, although not necessarily identical for these three models, is higher, and the number of cores is smaller, than suggested by Hill & Marty and Cassidy et al. models. The optimal speedup, although different for each of these three models, is consistently lower than the one suggested by Hill & Marty and Cassidy et al. models.

As can be seen in Figure 6, the optimal symmetrical multicore speedup is affected by $f$ as well as by synchronization and communication overhead. We discuss this effect in Section 4.1 below.

## 2.4 Asymmetric multicore performance model

The speedup function of the asymmetric multicore (relative to using one single BCE core as defined in [2]) can be presented by modifying (15) as follows:

$$Speedup_{asym} = \frac{Perf_{seq}(r)}{(1-f) + \frac{f}{s} + \frac{T_C}{n_c T_1} + \frac{T_S}{T_1}} \quad (17)$$

where $s$ is the parallel multicore speedup factor, as follows:

$$s = \frac{Perf_{par}}{Perf_{seq}} = \frac{\sqrt{r} + n - r}{\sqrt{r}} \quad (18)$$

Substituting (18) in (17), we can rewrite the asymmetric multicore speedup as follows:

$$Speedup_{asym} = \quad (19)$$
$$= \frac{\sqrt{r}}{(1-f) + \frac{f\sqrt{r}}{\sqrt{r} + n - r} + \frac{f_1}{n - r + 1} + f_2}$$

where $n_c = n - r + 1$ is the number of cores in asymmetric multicore. Substituting $f_1 = f_2 = 0$ in (19) yields Hill and Marty asymmetric multicore model [2].

Figure 7 shows the speedup of the asymmetric multicore using our model (red), the Hill and Marty model (blue), and the Eyerman and Eeckhout model [7] (black). Cassidy et al. and Gunther et al. models are not included as they do not address asymmetric architectures. For modeling purposes we use the same assumptions as in the case of the symmetric multicore (Figure 6).

Similarly to the symmetric multicore, when parallelism is low (f=0.5), all models behave similarly. The difference among the models becomes more apparent as $f$ grows. The optimal speedup is consistently lower under our model than according to Hill and Marty or Eyerman and Eeckhout models. The optimal size of the sequential core $r$ is consistently larger in our model than under Hill and Marty or Eyerman and Eeckhout models, and consequently the optimal number of cores according to our model is smaller.

Similarly to the symmetric multicore, we discuss the effects of the parallelism as well as the synchronization and communication overhead on the optimal asymmetrical speedup in Section 4.1 below.

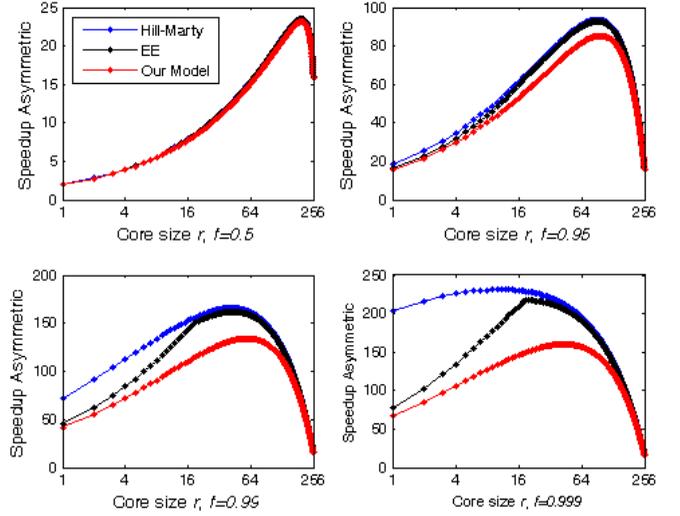

Figure 7. Speedup of the asymmetric multicore vs. sequential core size $r$

## 3 SPEEDUP SIMULATION

We simulate a symmetric multicore with a variable core size $r$. We simulate three highly parallelizable workloads with $f \to 1$. The goal of the simulation is to validate our analytic model.

### 3.1 Simulator

A multicore architectural simulator has been constructed for this study. It models a symmetric multicore with a parameterized number of cores, each equipped with a private memory, and a last-level shared memory. After the sequential portion of a workload is executed, the relevant data is uploaded from the private memory to the last-level shared memory. Before the execution of the parallelizable portion begins, the relevant input data are downloaded from the last level shared memory to the private memories of the individual cores. After the parallelizable portion is completed, the output data are uploaded from the private memories of the individual cores to the last level shared memory. Inter-core communications are performed through a simple switch with a number of predefined permutations, including butterfly (for FFT) and 2D mesh. We use a barrier synchronization model to allow for a parallel (easily scalable) inter-core data communication. Instructions, including memory access, take a predefined number of cycles to execute.

This approach allows capturing dynamic and transient time-dependent effects of the fine-grained data synchronization and inter-core communication. The simulator further exposes dynamic variability within each workload, which is not captured by our analytical model.

### 3.2 Workload

The workloads selected for performance analysis are the same ones discussed above in Sect.2.1:
- 256-option pairs Black-Sholes pricing (single precision floating point)
- 256-point Fast Fourier Transform (FFT, Radix 2, single precision floating point)

- Dense Matrix Multiplication of two 16×16 matrices (single precision floating point)

These workloads are significant because they exhibit three distinct types of synchronization and connectivity intensities (Figure 5). For dense matrix multiplication and FFT, we used optimized implementations outlined in [19]. For Black-Scholes, we used a direct implementation of the formulation in [31]. Note that in all three cases, sequential-to-parallel synchronization has been used mainly for uploading the data to and downloading the results from the multicore memory, and there are no serial code segments ($f \to 1$).

### 3.3 Results

The purpose of the simulation is to validate the analytic results obtained in Sect. 2. We focus on speedup vs. core size $r$. The simulation parameters are summarized in Table 1; serial execution times in the table are used to compute the speedup as per (16).

Simulation results are presented in Figure 8. In addition to actual speedup (red, blue and green curves), we plotted the theoretical speedup in black (as per our analytical model (16)). While we assumed $f_2 = 0.01$ in the theoretical model, $f_2$ values found by simulation are given in Table 2, and $f_1(r)$ simulated values are presented in Figure 9.

TABLE 1
SIMULATION PARAMETERS

| Parameter | Value |
|---|---|
| Task Size $N$ | 256 |
| Multicore Size, BCE | $N$ |
| $N$ FFT serial execution time | $5N \log_2 N$ |
| $\sqrt{N} \times \sqrt{N}$ DMM serial execution time | $2N^{1.5}$ |
| $N$ Black-Scholes serial execution time | $560 \cdot N$ |
| Number of cores $n$ | 1 through $N$ |
| Core size $r$ | $N$ through 1 |
| Arithmetic operation (IEEE Single Point Floating Point Multiplication/Addition) | 1 cycle |
| Interconnect distance | 1 |

Differences in simulation behavior are the result of workload properties. Black-Sholes option pricing is an 'embarrassingly parallel' workload with connectivity intensity $f_1 = 0$, as there are no inter-core communications. The simulated synchronization intensity (0.014) is somewhat higher than assumed (0.01), and consequently the simulated speedup is somewhat lower than the predicted speedup.

TABLE 2
SIMULATED SYNCHRONIZATION INTENSITY VALUES

| Workload | Synchronization Intensity |
|---|---|
| Black-Sholes option pricing | 0.014 |
| Fast Fourier Transform | 0.05 |
| Dense Matrix Multiplication | 0.094 |

Both simulated synchronization and connectivity intensities of FFT are higher than the assumed value (Table 2 and Figure 9). As a result, the maximum simulated speedup is more than three times lower than the predicted speedup.

Dense matrix multiplication achieves the lowest speedup. This is because both simulated synchronization and connectivity intensities of dense matrix multiplication are significantly higher than assumed in the theoretical model (Table 2 and Figure 9).

## 4 DISCUSSION

### 4.1 Speedup limitation, results and implications

Amdahl's law states that the speedup on $n_c$ processors is governed by the following:

$$Speedup(f, n_c) = \frac{1}{1 - f + \frac{f}{n_c}} \quad (20)$$

In this work we find that sequential-to-parallel data synchronization and inter-core communication affect Amdahl's law in the following way (*cf.* Sect. 2.3):

$$Speedup(f, n_c) = \frac{1}{1 - f + \frac{f}{n_c} + \frac{f_1(n_c)}{n_c} + f_2(n_c)} \quad (21)$$

It is apparent that for lower $f$ (parallelizable fraction of the program) our model behaves similarly to Hill and Marty model, which means that at low levels of $f$ ignoring the effects of sequential-to-parallel data synchronization and inter-core communication is acceptable. But as $f$ approaches 1, these effects become more predominant, considerably changing the overall picture. Specifically, both Hill and Marty and Cassidy models reach a different conclusion as to the optimal configuration.

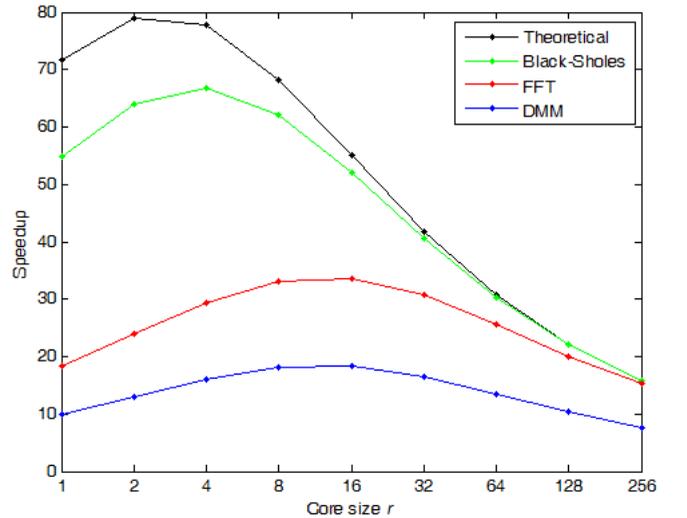

Figure 8. Simulation results: Speedup vs. core size

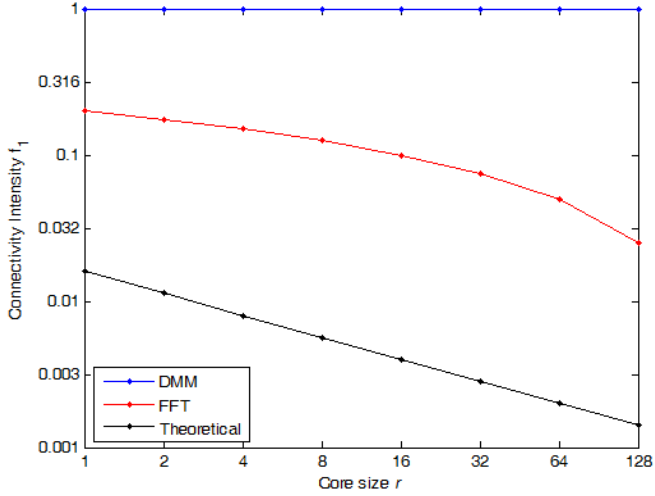
Figure 9. Simulation results: Connectivity intensity vs. core size

For symmetrical multicore, the best configuration yielded by Hill and Marty and Cassidy models is the maximum number of smallest cores, while in our model the optimal speedup is achieved with a smaller number of larger cores. For asymmetrical multicore, the optimal size of the large core is also larger (and therefore the number of small cores is lower) than the one yielded by Hill and Marty model. Consequently, the following results may be formulated:

**Result 1**: The impact of sequential-to-parallel synchronization and inter-core communication on multicore performance grows with parallelism.

**Result 2**: Even for highly parallelizable programs, a smaller number of larger cores seems to outperform a larger number of smaller ones, contrary to Hill and Marty model and contrary to the implication of Amdahl's law. This is similar to the result reached by Eyerman and Eeckhout [7], but for a different reason. In Eyerman and Eeckhout case, execution of critical sections requires faster and therefore larger cores. In our research, the reason for performance penalty is sequential-to-parallel and inter-core communications, which cannot be sped up simply by adding raw processing power. The model shows that a smaller number of larger cores achieved better performance, possibly thanks to less contention in the communication network, less contenders for shared resources, shorter distances (fewer number of hops) etc.

Assume that the synchronization intensity $f_2 = O(N^q) = O(n_c^q) = f_2' \cdot n_c^q$, where $f_2'$ is an application-specific constant (following [10], we assume for mathematical simplicity that workload is scaled to the multicore size, i.e. $N = n_c$). For example, for Black-Scholes option pricing, $q = 0$; for dense matrix multiplication, $q = -1$. If $q < 0$, the limit of both symmetric multicore speedup (16) and asymmetric multicore speedup (19), as $n_c$ grows, is:

$$\lim_{n \to \infty} Speedup = \frac{1}{1-f} \quad (22)$$

That is, accounting for data synchronization between the sequential and the parallel portions of a program does not change the main premise of Amdahl's law (20). This is the case for FFT and dense matrix multiplication considered above. If however $q = 0$ (as in Black and Scholes option pricing), then

$$\lim_{n \to \infty} Speedup = \frac{1}{1 - f + f_2'} \quad (23)$$

If $q > 0$, the speedup can diminish considerably by the sequential-to-parallel data synchronization delay as $n_c$ becomes very large. This problem is well known to hardware accelerator designers [20]. Similarly, given the connectivity intensity $f_1 = O(n_c^p)$ where $p > 1$, the speedup can be considerably diminished by the inter-core data communication delay as $n$ becomes very large.

In general, given the synchronization intensity $f_2 = O(n_c^q)$ and connectivity intensity $f_1 = O(n_c^p) = f_1' \cdot n_c^p$, where $f_1'$ is an application-specific constant, the speedup is limited as follows:

$$\lim_{n \to \infty} Speedup = \begin{cases} \dfrac{1}{1-f} & p < 1, q < 0 \\ \dfrac{1}{1 - f + f_1' + f_2'} & p = 1, q = 0 \\ 0 & p > 1, q > 0 \end{cases} \quad (24)$$

This is similar to the result reached by Eyerman and Eeckhout [7] who augmented Amdahl's law by postulating that in addition to the sequential part, multicore speedup is limited also by the synchronization of parallel threads.

Figure 10 and Figure 11 show the maximum achievable multicore speedup vs. $f_2 = O(n_c^q)$ ($q$ is shown on the top axis) for symmetric and asymmetric multicore respectively. Since in our framework the only free variable is the core size $r$ (see (3), (4), (16) and (19)), we need to find the optimal core size $r$ that leads to the maximum speedup. For symmetric multicore, the optimal core size for Hill and Marty model is found as follows [3]:

$$\frac{\partial}{\partial r} Speedup_{sym} = \frac{\partial}{\partial r}\left[\frac{\sqrt{r}}{(1-f) + \frac{f \cdot r}{n}}\right] = 0 \quad (25)$$

$$\to r_{opt} = \frac{n(1-f)}{f}$$

For asymmetric multicore, the analytic solution does not exist [3], therefore we find the optimal core size and the maximum achievable speedup numerically. For our model, the optimal core size $r$ and the maximum achievable speedup for both symmetric and asymmetric multicores are found numerically.

For modeling purposes we assume total BCE count $n = 256$, $f_1 = 0$ and $f = 0.5, 0.95, 0.99$ and $0.999$. Horizontal lines correspond to the maximum speedup achieved by Hill and Marty model.

The other curves, following our model, demonstrate that achievable maximum speedup may be lower, depending on $f_2$, as predicted by (24): for $q < 0$, the speedup grows to eventually reach the upper bound set by Hill and Marty model; for $q > 0$, the speedup diminishes.

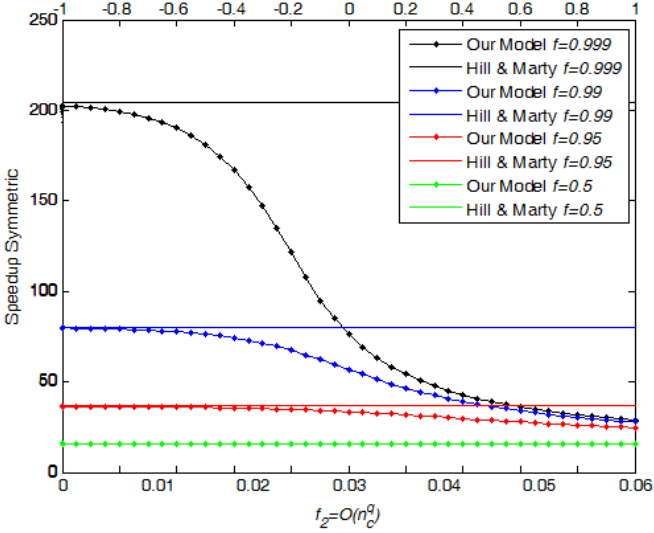

Figure 10. Maximum symmetric speedup vs. $f_2$

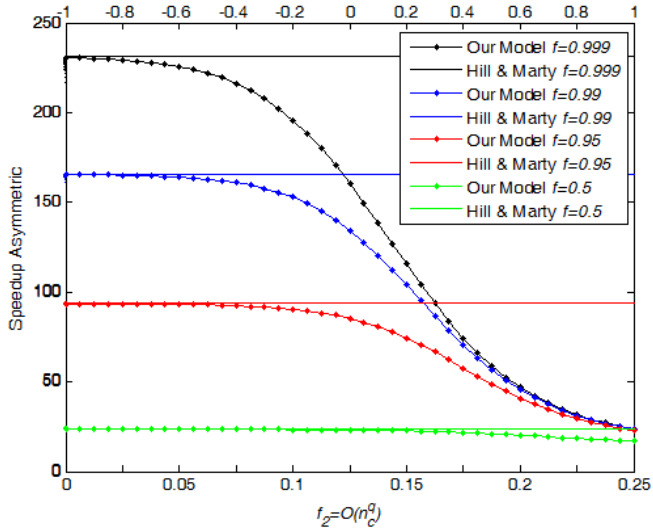

Figure 11. Maximum asymmetric speedup vs. $f_2$

Result 3: For tasks of high synchronization intensity (low arithmetic intensity), parallel execution may result in significantly lower speedup than predicted by Amdahl's law or Hill and Marty model.

Hill and Marty ([2], Implication 1) concluded: "Researchers should target increasing $f$ through architectural support, compiler techniques, programming model improvements, and so on". As our model reveals, the higher the parallelization becomes, the stronger the damping effects of sequential to parallel synchronization and inter-core communication on multicore performance. Therefore, we propose to amend Hill and Marty Implication 1 as follows:

**Implication 1**: Researchers should target increasing $f$ if $f$ is low. Beyond a certain point, targeting parallelism alone becomes less efficient. We find that this point lies around $f = 0.95$ for the symmetric multicores and $f = 0.99$ for the asymmetric multicore since around this point, the speedup predicted by our model begins to considerably deteriorate relative to the upper-bound speedup predicted by the Hill and Marty model for a practical range of $f_2$ values (0.01 ÷ 0.1) (see Figure 10 and Figure 11). Researchers should increasingly target reducing data dependencies among parallel program segments and decreasing synchronization intensity (increasing arithmetic intensity) of the parallel fractions along with improving parallelism.

A question arises if the effects of sequential-to-parallel data synchronization and inter-core communication can be overcome by proper architectural and software design. If a multicore architecture had no private memory (that is, the sequential core and the parallel multicore shared a common memory), then there would be no need in synchronization and/or inter-core communication – any data item could be reached by each core using simple shared memory access. Therefore such shared-memory architecture would be unaffected by those limiting factors. Obviously this may come at the penalty of a longer latency per each memory access. That latency can be further mitigated by introducing multithreading. This would require many threads per core, e.g. as discussed in [24].

The notion that sequential-to-parallel synchronization does not scale is not unequivocal. Consider the following example: sequential-to-parallel synchronization is done not directly between the individual memories of sequential and parallel cores, but through a higher level shared memory as noted in Sect. 2.1.1. Suppose such shared memory is implemented as a number of individual modules, connected to the memories of sequential and parallel cores through an interconnection network. Further assuming that there are no data dependencies among the modules of the shared memory (e.g. as in Black-Scholes but neither in FFT nor in dense matrix multiplication), sequential-to-parallel synchronization may scale by the number of modules of the shared memory. This however does not change the conclusions and implications of this work.

### 4.2 Synchronization and connectivity intensity-aware scheduler

Figure 12 shows the optimal symmetric core size $r$ achieved by our model vs. the parallelizable fraction of a program $f$ for $f_1 = O(n_c^{0.5})$ $f_1 = O(n_c^{0.75})$ and $f_1 = O(n_c)$. The optimal symmetric core size according to Hill and Marty model is shown as well. Figure 13 shows the optimal symmetric core size $r$ achieved by our model vs. the parallelizable fraction of a program $f$ for $f_2 = O(n_c^{0.5})$ $f_2 = O(n_c^{0.75})$ and $f_2 = O(n_c)$. The optimal symmetric core size according to Hill and Marty model is shown as well.

Similarly to Sect. 4.1, the optimal core size $r$ is found analytically for Hill and Marty model of symmetric multicore, as in (25). For Hill and Marty model of asymmetric multicore as well as for our models, the optimal core size $r$ is determined numerically.

The results are consistent with the ensuing analysis: as

$f$ approaches 1, the optimal core size approaches zero (and the number of cores approaches infinity) in Hill and Marty model (consistent with Amdahl's law), while in our model, the optimal core size converges to a positive value, depending on $f_1$ (Figure 12) or $f_2$ (Figure 13).

As discussed above, both $f_1$ and $f_2$ are application dependent. Therefore it would be beneficial to develop a scheduler that is $f_1$-aware, so that for applications with high connectivity intensity, the workload should be assigned to fewer cores, even if the parallelization fraction $f$ is very close to 1. For applications with lower connectivity intensity, the workload should be assigned in accordance with $f$ (the higher the $f$, the more cores that should be assigned work), as implied by Amdahl's law.

Similarly, it would be advantageous to develop a $f_2$-aware scheduler, as for high synchronization intensity (low arithmetic intensity) applications, it could be more efficient to keep the parallelizable portion of the program assigned to the sequential core, to spare the sequential-to-parallel synchronization delay, even for applications with $f$ very close to 1. Dense matrix multiplication is an example of such an application. Alternatively, for low synchronization intensity (high arithmetic intensity) applications, the parallelizable portion of a program should be assigned to the parallel cores in accordance with $f$, as implied by Amdahl's law.

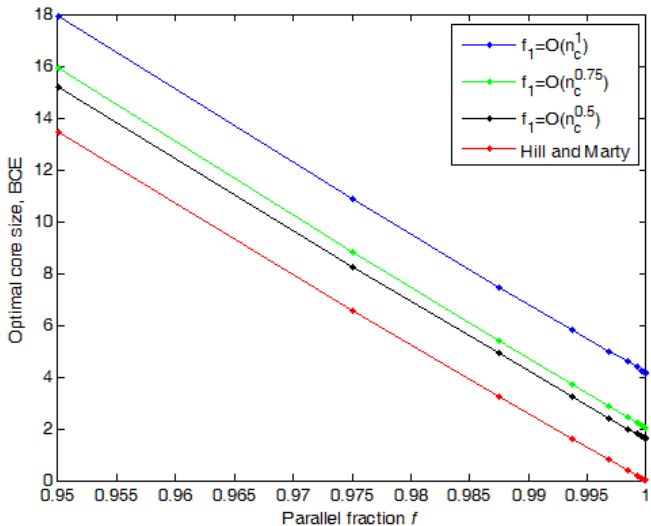

Figure 12. Optimal symmetric core size vs. $f$ for various values of $f_1$

## 5 SUMMARY

We have presented a model of parallel symmetric and asymmetric multicore performance taking into account the effects of sequential-to-parallel data synchronization and inter-core data communication. We performed a comparative analysis of our multicore performance model relative to a number of existing analytic multicore performance speedup models. We have demonstrated that even for highly parallelizable algorithms, the scalability is limited by delays introduced by sequential-to-parallel data synchronization and inter-core data communication. The maximum performance speedup achieved by multicore can be significantly lower than predicted by Amdahl's law, due to properties of each algorithm (synchronization and connectivity intensities). Consequently a highly parallelizable yet highly synchronization- and connectivity-intensive algorithm might be more efficiently implemented by a sequential (or single) core rather than by a parallel multicore. To improve the scalability and performance speedup of a multicore, it is as important to address the synchronization and connectivity intensities of parallel algorithms as their parallelization factor.

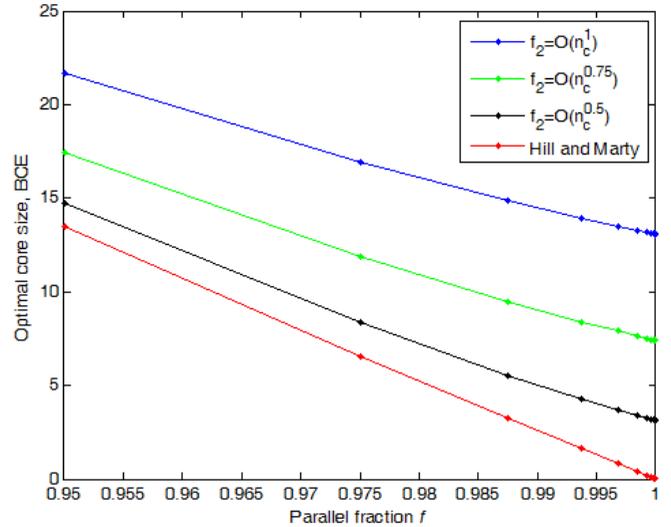

Figure 13. Optimal symmetric core size vs. $f$ for various values of $f_2$

## 6 REFERENCES

[1] G. Amdahl. "Validity of the single processor approach to achieving large scale computing capability." In Proc. AFIPS Spring Joint Computer Conf., 1967, pp. 483-485.

[2] M.D. Hill, M.R. Marty, "Amdahl's law in the multicore era", IEEE Computer 41 (7) (July 2008) 33–38.

[3] Erlin Yao, Yungang Bao, Mingyu Chen, "Extending Amdahl's Law in the Multicore Era", ACM SIGMETRICS Performance Evaluation, Volume 37 Issue 2, September 2009

[4] T. Oh, H. Lee, K. Lee, and S. Cho, "An Analytical Model to Study Optimal Area Breakdown between Cores and Caches in a Chip Multiprocessor," in IEEE Computer Society Annual Symposium on VLSI. IEEE Computer Society, 2009, pp. 181–186

[5] A. S. Cassidy and A. G. Andreou, "Analytical methods for the design and optimization of chip-multiprocessor architectures", 43rd Annual Conference on Information Sciences and Systems, 2009

[6] A. Cassidy, Kai Yu, Haolang Zhou and A. Andreou, "A High-Level Analytical Model for Application Specific CMP Design Exploration", Design, Automation & Test in Europe Conference & Exhibition (DATE), 2011

[7] A. Cassidy and A. Andreou, "Beyond Amdahl Law - An objective function that links performance gains to delay and energy", IEEE Transactions on Computers, vol. 61, no. 8, pp. 1110-1126, Aug 2012.

[8] Stijn Eyerman, Lieven Eeckhout, "Modeling Critical Sections in Amdahl's Law and its Implications for Multicore Design" in ISCA' 10: Proceedings of the 37th Annual International Symposium on Computer Architecture. New York, NY, USA: ACM, 2010, pp. 362–370.

[9] E. S. Chung, P. A. Milder, J. C. Hoe, and K. Mai. Single-chip heterogeneous computing: Does the future include custom logic, FPGAs, and GPGPUs? In MICRO-43, 2010.

[10] N. Gunther, S. Subramanyam, S. Parvu, "A Methodology for Optimizing Multithreaded System Scalability on Multi-Cores",

http://arxiv.org/abs/1105.4301

[11] Brian Rogers, Anil Krishna, Gordon Bell, Ken Vu, Xiaowei Jiang, Yan Solihin, "Scaling the Bandwidth Wall: Challenges in and Avenues for CMP Scaling". In ISCA '09: Proceedings of the 36th annual international symposium on Computer architecture, pages 371–382, New York, NY, USA, 2009. ACM

[12] G. Loh, "The Cost of Uncore in Throughput-Oriented Many-Core Processors", the Workshop on Architectures and Languages for Throughput Applications (ALTA), June 2008

[13] C Bienia, S Kumar, J Pal Singh, K Li, "The PARSEC benchmark suite: characterization and architectural implications", Proceedings of the 17th international conference on Parallel architectures and compilation techniques, ACM NY, NY, USA 2008

[14] S. Przybylski, M. Horowitz, and J. Hennessy, "Characteristics of performance-optimal multi-level cache hierarchies," Proceedings of the 16th Annual International Symposium on Computer Architecture (ISCA 89), pp. 114–121, Apr 1989.

[15] http://forums.nvidia.com/index.php?showtopic=224851

[16] M.J. Pertel, Mesh Distance Formulae, Technical Report, California Institute of Technology (CaltechCSTR:1992.cs-tr-92-05).

[17] T. Morad, Weiser U, Kolodny A, Valero M., Ayguade E., "Performance, power efficiency and scalability of asymmetric cluster chip multiprocessors", Computer Architecture Letters, Jan.-June 2006, Volume 5, Issue 1, pages 14 – 17.

[18] X.-H. Sun, Y. Chen, "Reevaluating Amdahl's law in the multicore era", Journal of Parallel and Distributed Computing, Volume 70, Issue 2, February 2010, Pages 183–188

[19] M.J. Quinn, "Designing Efficient Algorithms for Parallel Computers", McGraw-Hill, 1987, page 125.

[20] David Luebke, "General-purpose computation on graphics hardware", Workshop, SIGGRAPH, 2004

[21] M. Suleman, O. Mutlu, M. Qureshi, Y. Patt, "Accelerating critical section execution with asymmetric multi-core architectures", Proceedings of the International Conference on Architectural Support for Programming Languages and Operating Systems (ASPLOS), pages 253–264,Mar.2009.

[22] S. Kamil, C. Chan, L. Oliker,, J. Shalf, S. Williams, "An Auto-Tuning Framework for Parallel Multicore Stencil Computations", IEEE International Symposium on Parallel & Distributed Processing (IPDPS) 2010, pages 1-12.

[23] X.-H. Sun, L. Ni, Another view on parallel speedup, in: Proc. of IEEE Supercomputing'90,1990.

[24] Z. Guz, O. Itzhak, I. Keidar, A. Kolodny, A. Mendelson, U. Weiser, "Threads vs. Caches: modeling the behavior of parallel workloads", 2010 IEEE International Conference on Computer Design (ICCD), Oct. 2010, Pages: 274-281.

[25] X. Wen, U. Vishkin, "PRAM on chip: First commitment to silicon", Proceedings of the nineteenth annual ACM symposium on Parallel algorithms and architectures, SPAA 2007, pages 301-302.

[26] Samuel Williams, David Patterson, Leonid Oliker, John Shalf, Katherine Yelick, "The roofline model: A pedagogical tool for auto-tuning kernels on multicore architectures," Hot Chips 20, 2008.

[27] D. Pham et al., The Design and Implementation of a First-Generation Cell Processor, ISSCC Dig. Tech. Papers, Paper 10.2, 184-185, February, 2005.

[28] M. Taylor et al, "The Raw microprocessor: a computational fabric for software circuits and general-purpose programs", IEEE Micro, Volume 22, Issue 2, APRIL 2002, Pages 23-25

[29] D. Fan, " Godson-T: An Efficient Many-Core Architecture for Parallel Program Executions", IEEE Micro, Volume 24, Issue 9, 2009, Pages 1063-1071

[30] D. Wentzlaff et al, "On-Chip Interconnection Architecture of the Tile Processor", IEEE Micro, Volume 27, Issue 5, OCTOBER 2007, Pages 15-31.

[31] F. Black and M. Scholes, "The pricing of options and corporate liabilities," Journal of Political Economy, 81 (1973), pp. 637–654, 1973.

[32] H. P. Flatt. "Performance of Parallel Processors." In Parallel Computing, Vol. 12, No. 1, 1989, pp. 1-20.